\documentclass[10pt,a4paper]{article}
\usepackage[english]{babel}
\usepackage{mathpazo}
\usepackage{a4wide}
\usepackage{bm,amsmath,amssymb,accents}
\usepackage{graphicx}
\usepackage{subfigure}
\usepackage{units}
\usepackage{xspace}
\usepackage{multirow}
\usepackage{xcolor}
\definecolor{darkred}{rgb}{0.5,0,0}
\definecolor{darkblue}{rgb}{0,0,0.5}
\definecolor{firebrick}{rgb}{0.75,0.125,0.125}
\definecolor{darkgreen}{rgb}{0,0.5,0}
\usepackage[colorlinks=true,linkcolor=firebrick,citecolor=darkgreen,urlcolor=darkblue]{hyperref}

\def\HRule{\rule{\linewidth}{1mm}}
\def\gap#1#2{GAP-{#1}-{#2}}

\def\eq#1{\begin{equation}#1\end{equation}}
\def\al#1{\begin{align}#1\end{align}}
\def\d{{\rm d}}
\def\onefig{\textwidth}
\def\twofig{0.48\textwidth}

\def\p{\protect\raisebox{-0.4ex}{\protect\resizebox{!}{1.4ex}{$\mathcal{P}$}}}
\def\P{\mathcal{P}}
\newcommand{\EXP}[2][]{{{\mathbb E}_{#1}\left\{{#2}\right\}}}
\newcommand{\VAR}[2][]{{{\mathrm{Var}}_{#1}\left\{{#2}\right\}}}

\begin{document}

\thispagestyle{empty}
\vspace*{\stretch{1}}
\begin{flushright}
  \HRule
  \\[9mm]
  \large
  \gap{2009}{043}
  \\[7mm]
  {\bf\Huge Single Muon Response: Tracklength}
  \\[10mm]
  \includegraphics[height=35mm]{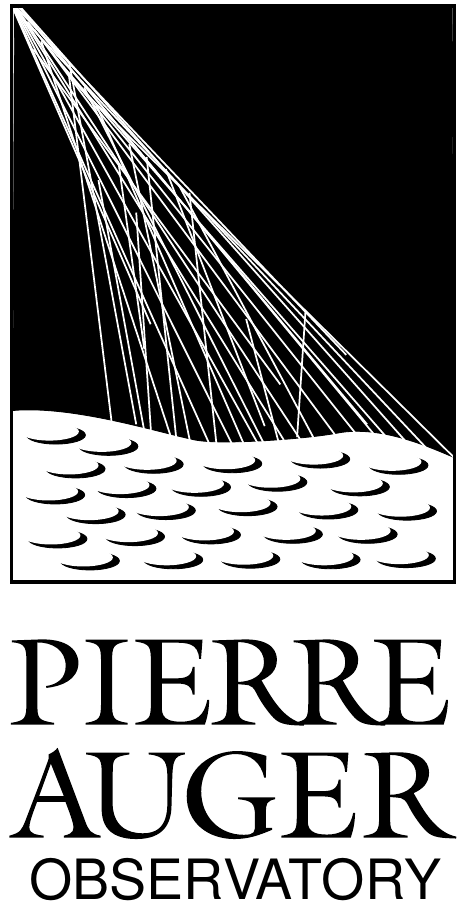} \hfill%
  \parbox[b]{13cm}{\begin{flushright}
    {\Large
      Bal\'azs K\'egl$^{\,a}$ and
      Darko Veberi\v{c}$^{\,b}$
    }
    \\[6mm]
    \parbox[b]{10cm}{\begin{flushright}
      $^{\,a}$ LAL, IN2P3 \& University Paris-Sud, France
      \\
      $^{\,b}$ University of Nova Gorica, Slovenia
    \end{flushright}}
    \\[10mm]
    March 2009
  \end{flushright}}
  \\[5mm]
  \HRule
\end{flushright}
\vspace*{\stretch{1}}
\begin{abstract}
In this note we aim to infer a model for
the response of a Pierre Auger water-Cherenkov detector to an ideal single
muon. The main goal of this analysis is to provide analytical support for muon
counting techniques. In this note we derive the probability distribution of the
muon tracklength as a function of the zenith angle of the muon.
\end{abstract}
\vspace*{\stretch{3}}
\clearpage

\section{Introduction}

This note is the first in a series of notes in which we aim to infer a model for
the response of a Pierre Auger water Cherenkov detector (\emph{tank} from now
on) to an ideal single muon. The single muon response is a subject that was
thoroughly explored in the early phase of the Auger collaboration
\cite{gap1996_008, gap1996_011, gap1997_004, gap1997_024, gap1997_026,
gap2001_018, gap2002_045, gap2002_063, gap2002_078, gap2003_054, gap2003_113,
gap2004_045, gap2005_010, gap2005_101, thesisDornic06, gap2009_038}. The main
purpose of these studies was to understand the mean muon response (more
precisely, the so-called muonic peak or the ``muon hump'') in order to define an
SD energy estimate. First, based on the muonic peak model, a calibration
procedure was designed for estimating the total signal in individual tanks. Then
the total signals were combined to compute one observable per shower, which was
finally calibrated to the energy estimate of the fluorescence detector (FD).

Since the SD energy estimate is calibrated to the FD energy estimate, the
procedure is relatively robust against systematic biases of the mean muon
response estimate as long as the biases are stable among different tanks and not
changing with time. The fluctuations of the muonic signal are also not too
important from the point of view of the energy estimation. On the other hand,
the goal of muon-counting \cite{gap2006_065, gap2007_060, gap2008_136,
gap2009_023} is to design a muon density estimator without the need for outside
calibration. Now biases due to the tank geometry and the energy-dependence of
the muonic response become important. Furthermore, since these techniques are
based on certain statistics of the individual muonic ``jumps'', understanding
the fluctuations of the muon response is of great importance. Our ultimate goal
is similar to the program outlined in \cite{gap2004_045}: obtain a full
parametrization of the muonic signal that can be used in a Monte Carlo Markov
chain \cite{RoCa04} reconstruction approach as well as for fine-tuning the muon
counting techniques. Beside this principal goal, we believe that this refined
model may also help to improve the SD energy estimate (mainly by decreasing the
statistical error on the individual shower estimates along the lines of
\cite{gap2005_054}). The obtained model may also serve as a basis for a toy
Monte-Carlo tank simulator that can be used to quickly generate a large number
of tank signals.

\subsection{Introduction to this note}

In a crude model, the total muonic signal is proportional to the tracklength of
the muon in the water tank, so the fluctuation of the tracklength at a given
zenith angle is the main source of the fluctuation of the total muonic
signal. In this note we derive the tracklength distribution $\p_L(L \mid \theta)$ of
a muon that crosses the tank at zenith angle $\theta$. Most of the formulas can
also be found in \cite{gap2003_113}. Our main contribution is that we give clean
formulas that can be used directly in a probabilistic generative model. We also
add one term for muons that enter and leave at the lateral of the tank. This
part of the signal, which was omitted by \cite{gap2003_113}, may be important
for inclined showers.

\section{The distribution of the muon tracklength}

In this section we derive the tracklength distribution
$\p_L(L \mid \theta)$ of a muon that crosses the tank zenith angle
$\theta$. To transform the quantities into unit-less variables, we
first apply the substitutions
\eq{
\ell = \frac{L}{2R}
\qquad\text{and}\qquad
h = \frac{H}{2R},
}
where $h=1/3$ for an Auger tank with height $H=\unit[1.2]{m}$ and
radius $R=\unit[1.8]{m}$. From now on we will only derive the
distribution $\p(\ell \mid \theta)$ as a function of the unit-less variable
$\ell$. The original distribution in $L$ can be recovered by the
transformation
\eq{
\p_L(L \mid \theta) = \frac{1}{2R} \p(L/2R \mid \theta).
}

\begin{figure}[!ht]
\centering
\includegraphics[width=\twofig]{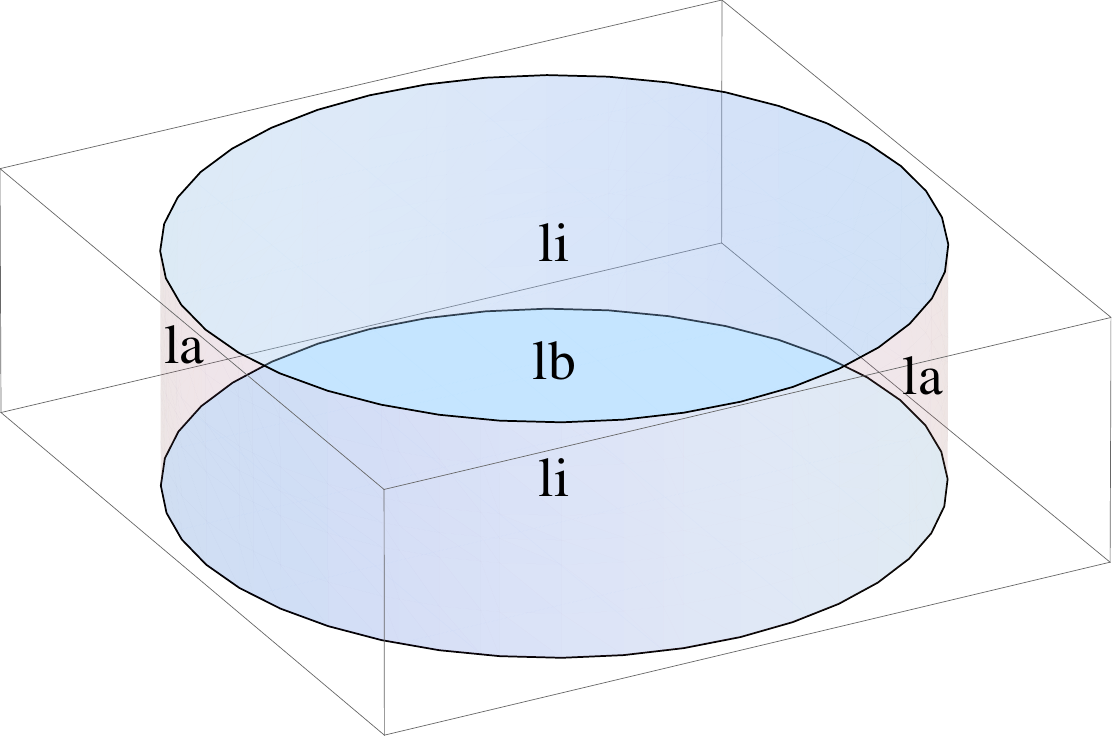}
\caption{Projection of the Auger tank along the incoming muon direction at
$\theta=50^\circ$. The three components of the tracklength
distribution $\p(\ell \mid 50^\circ)$ are visible: the \emph{lid--base}
(lb), the \emph{lid--lateral} (li), and the \emph{lateral--lateral} (la)
terms.}
\label{f:Tank60ColorLabeled}
\end{figure}

The distribution can be decomposed into three components: the
\emph{lid--base} (lb) term of the form
$\p_\text{lb}(\ell \mid \theta)\,\P(\text{lb} \mid \theta)$ representing muons
entering at the lid of the tank and leaving at the base, the
\emph{lid--lateral} (li) term 
$\p_\text{li}(\ell \mid \theta)\,\P(\text{li} \mid \theta)$
representing muons entering at the lid and leaving at the lateral (or,
symmetrically, entering at the lateral and leaving at the base), and
the \emph{lateral--lateral} (la) term
$\p_\text{la}(\ell \mid \theta)\,\P(\text{la} \mid \theta)$ representing muons
entering and leaving at the lateral (for these regions see projected
tank schematic in Fig.~\ref{f:Tank60ColorLabeled}). We have thus
\eq{
\p(\ell \mid \theta) =
  \p_\text{lb}(\ell \mid \theta)\,\P(\text{lb} \mid \theta) +
  \p_\text{li}(\ell \mid \theta)\,\P(\text{li} \mid \theta) +
  \p_\text{la}(\ell \mid \theta)\,\P(\text{la} \mid \theta).
\label{prob}
}

\subsection{Case probabilities}

\begin{figure}[!ht]
\centering
\subfigure[relative areas]{
\includegraphics[width=\twofig]{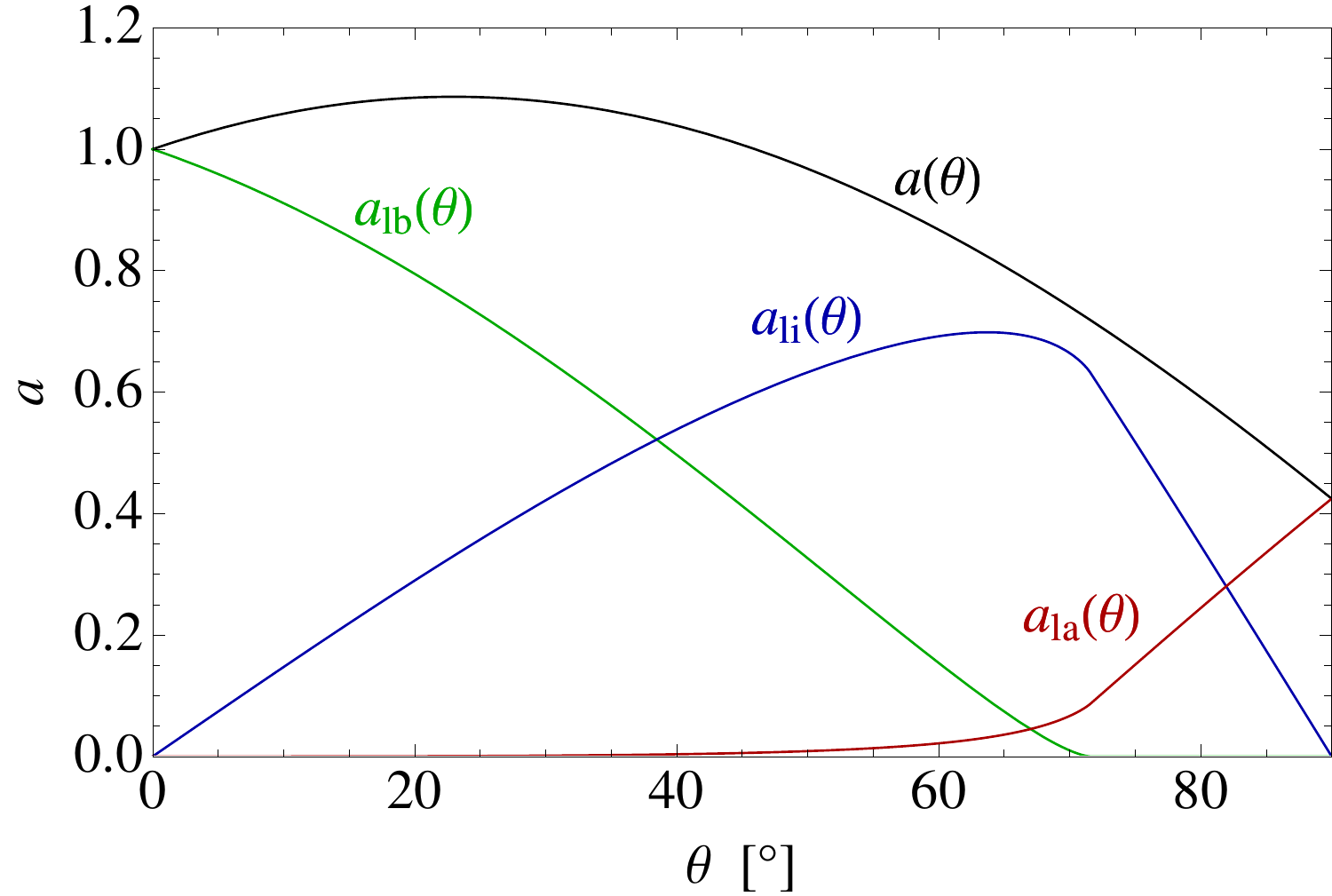}
}
\subfigure[case probabilities]{
\includegraphics[width=\twofig]{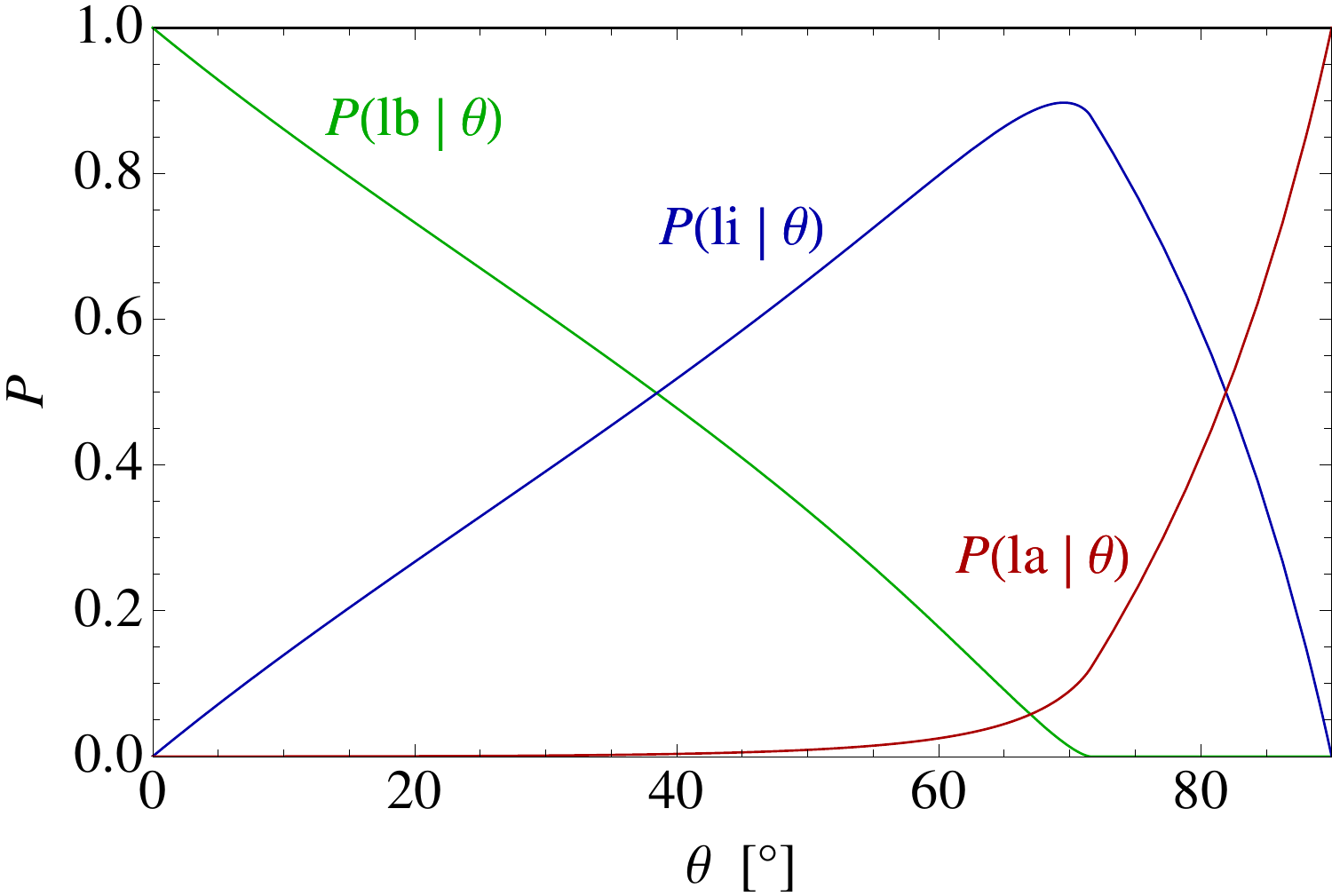}
}
\caption{(a) The relative partial areas $a_\text{lb}(\theta)$,
$a_\text{li}(\theta)$, $a_\text{la}(\theta)$, and their sum
$a(\theta)$, and (b) the case probabilities $\P(\text{lb} \mid \theta)$,
$\P(\text{li} \mid \theta)$, and $\P(\text{la} \mid \theta)$ as a function of the
zenith angle $\theta$.}
\label{f:TankGeometryII}
\end{figure}

We proceed by first computing the case probabilities $\P(\text{lb} \mid \theta)$,
$\P(\text{li} \mid \theta)$, and $\P(\text{la} \mid \theta)$. First we need the area
$A(\theta)$ of the tank projection for zenith angle $\theta$,
\eq{
A(\theta) = \pi R^2 |\cos\theta| + 2RH\sin\theta.
}
Note that the absolute value is used in order to render the equations
useful also for cases when muons are entering the tank with zenith
angle $\theta$ larger than $90^\circ$ (i.e., useful for albedo
studies). Measuring area in units of the top surface $A_\text{t}=\pi
R^2$ and using the unit-less $h$, we can define
\eq{
a(\theta) =
  \frac{A(\theta)}{A_\text{t}} =
  |\cos\theta|+\tfrac{4}{\pi}h\sin\theta =
  \tfrac{2}{\pi}|\cos\theta|(\tfrac{\pi}{2}+2u),
}
where
\eq{
u = h|\tan\theta|.
}
Measuring the partial areas (see Fig.~\ref{f:TankGeometryII}(a)) in the same
relative unit $a_\text{xx}=A_\text{xx}/A_\text{t}$, we obtain for the 
lid--base part
\al{
a_\text{lb}(\theta) & =
  \begin{cases}
    \tfrac{2}{\pi}\left(
      |\cos\theta|\arccos(h\tan\theta) -
      h\sin\theta\sqrt{1-(h\tan\theta)^2}
    \right) 
    & \text{ if $|\tan\theta|\leqslant1/h$,}
  \\[4mm]
    0 & \text{ otherwise,}
  \end{cases}
\\
  & =
  \tfrac{2}{\pi}|\cos\theta|
  \begin{cases}
    \arccos u - u \sqrt{1-u^2} & \text{ if $u \leqslant 1$,}
  \\[4mm]
    0 & \text{ otherwise,}
  \end{cases}
}
for the lid--lateral part
\al{
a_\text{li}(\theta) & =
  2 \big(|\cos\theta| - a_\text{lb}(\theta)\big) =
\\
  & = \tfrac{4}{\pi}|\cos\theta|
  \begin{cases}
    \arcsin u + u \sqrt{1-u^2} & \text{ if $u\leqslant1$,}
  \\[4mm]
    \tfrac{\pi}{2} & \text{ otherwise,}
  \end{cases}
}
and for the lateral--lateral part
\al{
a_\text{la}(\theta) & =
  \tfrac{4}{\pi}h\sin\theta - |\cos\theta| + a_\text{lb}(\theta) =
\\
  & = \tfrac{2}{\pi}|\cos\theta|
  \begin{cases}
    2u - \arcsin u - u \sqrt{1-u^2} & \text{ if $u \leqslant 1$,}
  \\[4mm]
    2u - \tfrac{\pi}{2} & \text{ otherwise.}
  \end{cases}
}
The case probabilities (see Fig.~\ref{f:TankGeometryII}(b)) are then
given by
\eq{
\P(\text{lb} \mid \theta) = \frac{a_\text{lb}(\theta)}{a(\theta)},
\qquad
\P(\text{li} \mid \theta) = \frac{a_\text{li}(\theta)}{a(\theta)},
\quad\text{and}\quad
\P(\text{la} \mid \theta) = \frac{a_\text{la}(\theta)}{a(\theta)}.
}
Note that for the final case probabilities, the $2|\cos\theta|/\pi$
term cancels in this process.

\subsection{Probability distribution functions}

\begin{figure}[!ht]
\centering
\includegraphics[width=\onefig]{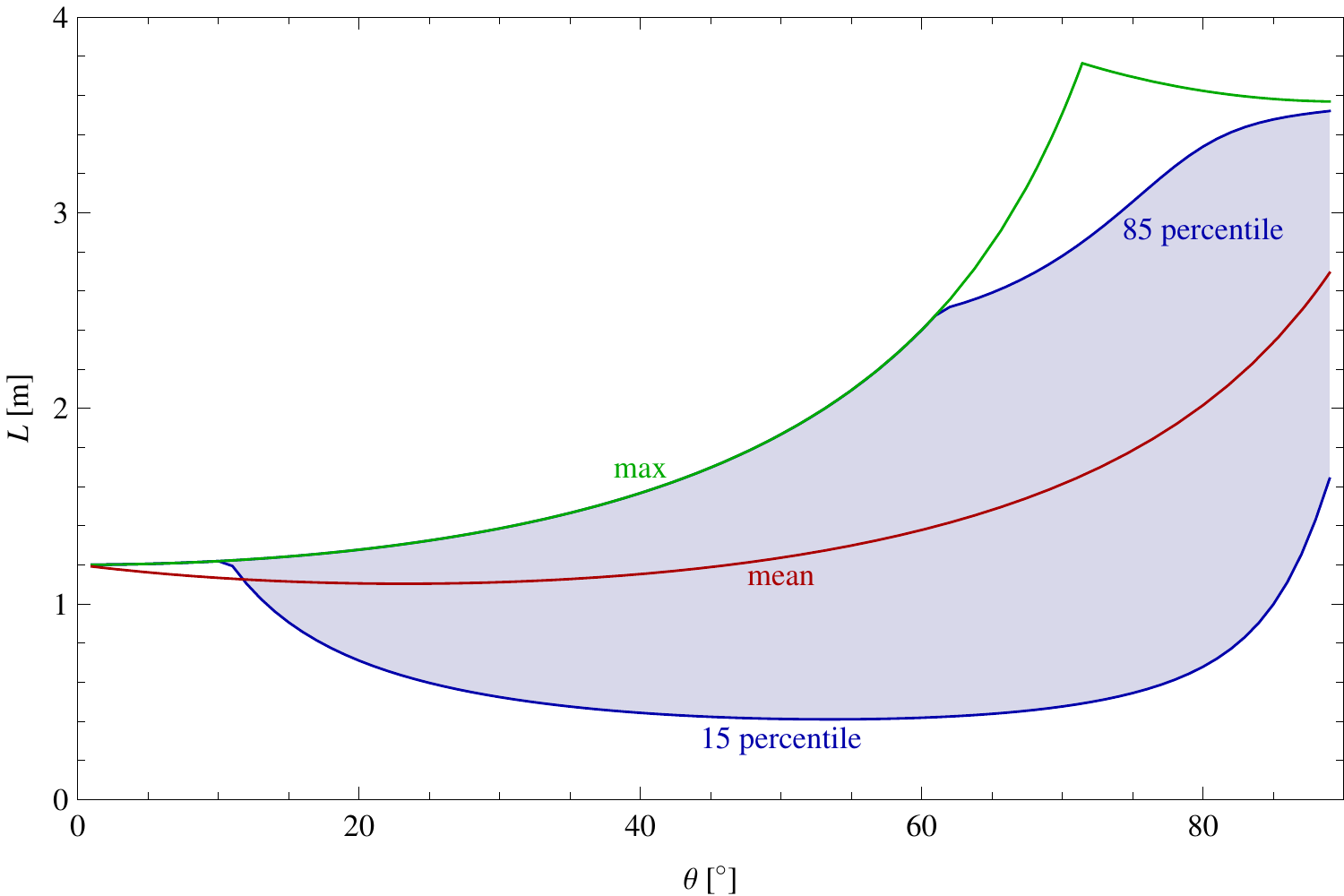}
\caption{The mean tracklength $\bar{L}(\theta) = \EXP{\p_L(L|\theta)}$ (red),
  the maximum tracklength $L_\text{max}(\theta) = 2R \ell_\text{max}(\theta)$
  (green), and the region that contains $70\%$ of the probability mass (blue) as
  a function of the zenith angle $\theta$.}
\label{f:lmax}
\end{figure}

The lid--base term $\p_\text{lb}(\ell \mid \theta)$ is a simple Dirac delta
\eq{
\p_\text{lb}(\ell \mid \theta) = \delta(\ell - \ell_\text{max}),
}
where $\ell_\text{max}(\theta)$ is the \emph{maximum tracklength}
(see Fig.~\ref{f:lmax}) defined as
\eq{
\ell_\text{max}(\theta) =
  \begin{cases}
    h/|\cos\theta| & \text{ if $|\tan\theta|\leqslant1/h$},
  \\
    1/\sin\theta & \text{ otherwise.}
  \end{cases}
}

Note that the case $1/\sin\theta$ is never in effect in the lid--base
term since the case probability $\P(\text{lb}|\theta)$ is zero in this
range. The lid--lateral and lateral--lateral terms are
\al{
\p_\text{li}(\ell \mid \theta) & =
  \frac{1}{a_\text{li}(\theta)}
  \begin{cases}
    \tfrac{4}{\pi}|\sin2\theta|\sqrt{1-(\ell\sin\theta)^2} &
      \text{ if $0\leqslant\ell\leqslant\ell_\text{max}(\theta)$,}
  \\[4mm]
    0 & \text{ otherwise,}
  \end{cases}
\\
\p_\text{la}(\ell \mid \theta) & =
  \frac{1}{a_\text{la}(\theta)}
  \begin{cases}
    \displaystyle
    \tfrac{4}{\pi}
    \frac{\ell\sin^3\theta(h-\ell|\cos\theta|)}
         {\sqrt{1-(\ell\sin\theta)^2}} &
      \text{ if $0\leqslant\ell\leqslant\ell_\text{max}(\theta)$,}
  \\[4mm]
    0 & \text{ otherwise,}
  \end{cases}
}
respectively.

\subsection{Cumulative distribution functions}

The cumulative distribution functions of the continuous terms are
\al{
\int_0^{\ell}\!\p_\text{li}(\ell' \mid \theta)\,\d\ell' &=
  \tfrac{4}{\pi}\frac{|\cos\theta|}{a_\text{li}(\theta)}
  \left[\arcsin(\ell\sin\theta) +
  \ell\sin\theta\sqrt{1-(\ell\sin\theta)^2}\right] =
\\
  & =
  \begin{cases}
  \displaystyle
  \frac{\arcsin v + v \sqrt{1-v^2}}{\arcsin u + u \sqrt{1-u^2}} 
    & \text{ if $u\leqslant1$,}
  \\[4mm]
  \displaystyle
  \frac{\arcsin v + v \sqrt{1-v^2}}{\pi/2} & \text{ otherwise,}
  \end{cases}
\\
\int_0^{\ell}\!\p_\text{la}(\ell' \mid \theta)\,\d\ell' &=
\\
  \tfrac{2}{\pi}\frac{|\cos\theta|}{a_\text{la}(\theta)}
  &\left[
    2h|\tan\theta|-\arcsin(\ell\sin\theta) -
    \left(2h|\tan\theta| - \ell\sin\theta
  \right)\sqrt{1 - (\ell\sin\theta)^2} \right] =
\\
  & =
  \begin{cases}
    \displaystyle
    \frac{2u - \arcsin v - (2u - v) \sqrt{1-v^2}}
          {2u - \arcsin u - u \sqrt{1-u^2}} 
    & \text{ if $u\leqslant1$,}
  \\[4mm]
    \displaystyle
    \frac{2u - \arcsin v - (2u - v) \sqrt{1-v^2}}{2u - \pi/2}
    & \text{ otherwise,}
  \end{cases}
}
where
\eq{
v = \ell\sin\theta,
}
and the full cumulative distribution is
\eq{
\int_0^{\ell}\!\p(\ell' \mid \theta)\,\d\ell' =
  \begin{cases}
  \displaystyle
  \frac{\arcsin v + 2u + (3v-2u) \sqrt{1-v^2}}{2u + \pi/2} 
    & \text{ if $v \leqslant \min(1,u)$,}
  \\[4mm]
  1 & \text{ otherwise.}
  \end{cases}
}

\subsection{Mean tracklength}

It has already been noted several times that the mean tracklength
$\bar{L}(\theta)$ (see Fig.~\ref{f:lmax}) for muons arriving with a specific
zenith angle $\theta$ is quite easy to obtain without integrating $L
\p_L(L \mid \theta)$. If we enlarge the puncturing track of a muon so that it has the
same cross section $\d A$ along the path, a small volume $\d V$ is
obtained. Adding up all such parallel tracks will eventually amount to the whole
volume of the tank,
\eq{
V=\int\!\d V=\int\!L\,\d A.
\label{volume}
}
The volume of the whole tank is $V=\pi R^2 H$ and the right integral
of Eq.~\eqref{volume} is in fact equal to the average tracklength
multiplied by the projected area $\int\!\d A=A(\theta)$. Therefore
$V=\bar{L}(\theta) A(\theta)$ and
\eq{
\bar{L}(\theta) =
  \EXP{\p_L(L \mid \theta)} =
  \frac{V}{A(\theta)} =
  \frac{H}{|\cos\theta|+\tfrac{2H}{\pi R}\sin\theta} =
  \frac{2Rh}{|\cos\theta|+\tfrac{4h}{\pi}\sin\theta}.
}

\section{Conclusion}

\begin{figure}[!ht]
\centering
\includegraphics[width=\onefig]{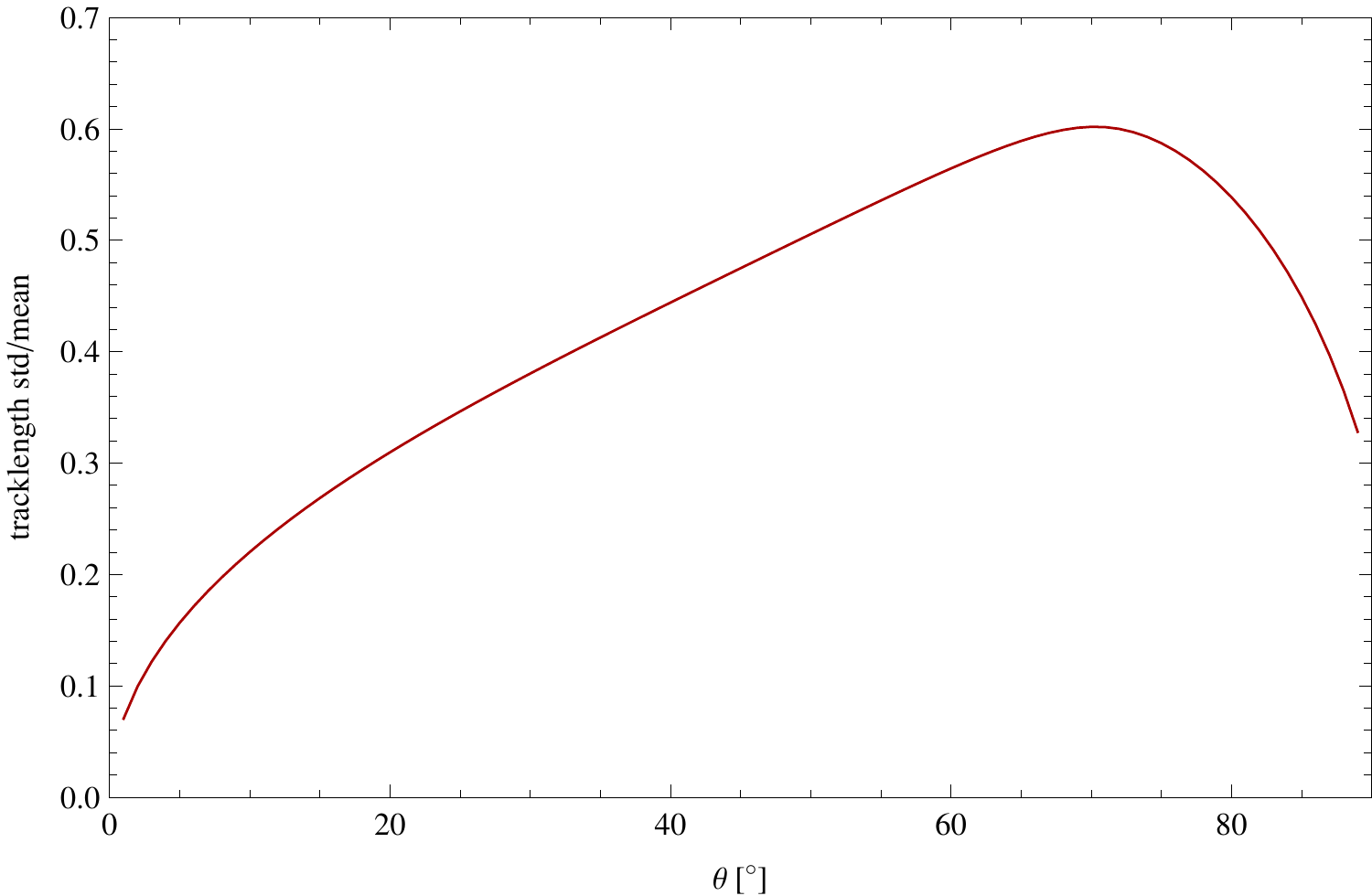}
\caption{The relative standard deviation
  $\sqrt{\VAR{\p_L(L \mid \theta)}}/\bar{L}(\theta)$ of the tracklength as a
  function of the zenith angle $\theta$.}
\label{f:rel}
\end{figure}

In this note we derived the probability distribution of the muon tracklength as
a function of its zenith angle $\theta$. The results show that the mean
tracklength is relatively stable in the $[0^\circ,60^\circ]$ interval, however
the fluctuations increase rapidly for $\theta > 10^\circ$ (see
Fig.~\ref{f:rel}). For inclined showers ($\theta > 60^\circ$), the mean
tracklength also increases by a factor of $2$.

\section*{Acknowledgments}

This work was supported by the ANR-07-JCJC-0052 grant of the French National
Research Agency.

\bibliographystyle{ieeetr}
\bibliography{balazs-darko}

\end{document}